\documentclass[aps,prb,reprint]{revtex4-1}
\usepackage{graphicx}
\usepackage{longtable}
\usepackage{amsmath}
\usepackage{hyperref}
\hypersetup{colorlinks,citecolor=blue,filecolor=black,linkcolor=blue,urlcolor=blue}

\begin{document}
	
	\title{The Zn vacancy as a polaronic hole trap in ZnO}
	
	\author{Y. K. Frodason}\email[E-mail: ]{ymirkf@fys.uio.no}
	\author{K. M. Johansen}
	\author{T. S. Bj\o rheim}
	\author{B. G. Svensson}
	\affiliation{University of Oslo, Centre for Materials Science and Nanotechnology, N-0318 Oslo, Norway}
	\author{A. Alkauskas}
	\affiliation{Center for Physical Sciences and Technology, Vilnius LT-10257, Lithuania}	
	
	\date{\today}
	
\begin{abstract}

	This work explores the Zn vacancy in ZnO using hybrid density functional theory calculations. The Zn vacancy is predicted to be an exceedingly deep polaronic acceptor that can bind a localized hole on each of the four nearest-neighbor O ions. The hole localization is accompanied by a distinct outward relaxation of the O ions, which leads to lower symmetry and reduced formation energy. Notably, we find that initial symmetry-breaking is required to capture this effect, which might explain the absence of polaronic hole localization in some previous hybrid density functional studies. We present a simple model to rationalize our findings with regard to the approximately equidistant thermodynamic charge-state transition levels. Furthermore, by employing a one-dimensional configuration coordinate model with parameters obtained from the hybrid density functional theory calculations, luminescence lineshapes were calculated. The results show that the isolated Zn vacancy is unlikely to be the origin of the commonly observed luminescence in the visible part of the emission spectrum from \textit{n}-type material, but rather the luminescence in the infrared region.

\end{abstract}
	
	\maketitle	
	
\section{\label{sec:introduction}Introduction}
	
	Although there is a plethora of density functional theory (DFT) studies addressing various aspects of intrinsic defects in ZnO, precise determination of properties like the formation energy and thermodynamic charge-state transition levels has been difficult, and the results scatter widely in the literature. This is especially evident for the Zn vacancy ($V_{\text{Zn}}$), where the reported thermodynamic charge-state transition levels spread over more than 2 eV \cite{Janotti2007,Oba2008,Clark2010,Demchenko2011,Johansen2015,Zhang2001,Kohan2000,Oba2001,Zhao2006,Janotti2009,Lany2007,Vidya2011,Sokol2007}. The main reason for this large variation can be divided into two categories \cite{Freysoldt2014,Clark2010}: (i) The choice of exchange-correlation functional, and (ii) The choice (or omission) of finite-size corrections. Despite this spread in results, DFT calculations have provided valuable insights into the properties of many defects in ZnO. The majority of studies conclude that $V_{\text{Zn}}$ is a deep acceptor---the dominant ``native'' acceptor-type defect---acting as as a compensating center in \textit{n}-type material \cite{Janotti2007,Oba2008}. However, theoretical studies that relied on semi-local and local density functionals, while providing valuable information, could not properly describe the localization of holes at $V_{\text{Zn}}$ as observed experimentally in, e.g., electron paramagnetic resonance (EPR) studies \cite{Galland1970,Galland1974,Kappers2008,Evans2008}.
	
	By employing the Heyd-Scuseria-Ernzerhof (HSE) hybrid functional \cite{Perdew1996,Heyd2003}, which intermixes a portion of screened Hartree-Fock (HF) exchange with the standard GGA-PBE exchange-correlation functional, we are able to capture the hole localization at $V_{\text{Zn}}$, which drastically modifies its properties. Furthermore, by using a one-dimensional configuration coordinate model \cite{Alkauskas2012}, defect luminescence lineshapes and positions for all optical transitions involving $V_{\text{Zn}}$ and the band edges are calculated. The results show that the isolated $V_{\text{Zn}}$ is unlikely to be the origin of the luminescence in the visible part of the emission spectrum from \textit{n}-type material.
	
	This paper is organized as follows. In Section \ref{sec:theoretical_framework}, we present computational details, outline how the various quantities are calculated and elaborate on the accuracy of the calculations. In Section \ref{sec:results}, the results are presented and discussed, including thermodynamics, electronic structure, a simple model for the energetics of $V_{\text{Zn}}$ and optical properties. Section \ref{sec:conclusion} concludes the paper.
		
\section{\label{sec:theoretical_framework}Theoretical framework}
		
\subsection{\label{sec:computational_details}Computational details}
	
	All calculations were performed using the projector-augmented wave (PAW) method \cite{Bloechl1994,Kresse1994,Kresse1999}, as implemented in the Vienna \textit{ab-initio} Simulation Package ({\scriptsize VASP}) \cite{Kresse1993,Kresse1996}, using a plane-wave energy cutoff of 500 eV. The Zn 3\textit{d}, 4\textit{s}, 4\textit{p}, and O 2\textit{s}, 2\textit{p} electrons were considered as valence electrons.  The $\alpha$-tuned HSE hybrid functional was used with a screening parameter \cite{Perdew1996,Heyd2003} of 0.2 \AA$^{-1}$, and the amount of exact exchange was set to $\alpha = 37.5 \%$. The resulting lattice parameters for wurtzite ZnO ($a=3.244$ \AA \ and $c=5.194$ \AA) and band gap (3.42 eV) are in excellent agreement with experimental data. Defect calculations were performed with a 96-atom-sized supercell by relaxing all ionic positions, but keeping its shape and volume fixed to that of the pristine supercell. Ionic optimization was performed until all forces were smaller than 5 meV/\AA, and the break condition for the electronic self consistent loop was set to $10^{-6}$ eV. Due to the periodic boundary conditions, defect wave functions may overlap causing an artificial dispersion of defect states. This may lead to an error in the defect formation energy for small supercells, particularly if a $\Gamma$-only \textit{k}-point sampling is used \cite{Makov1996,Freysoldt2014}. In this work, a special off-$\Gamma$ \textit{k}-point at $k=(\frac{1}{4},\frac{1}{4},\frac{1}{4})$ was employed in order to minimize this error within the bounds of computational cost. This setup was checked for several defects in ZnO against a 72 atom supercell with a 2x2x2 $\Gamma$-centered \textit{k}-mesh, yielding an average energy difference of merely 0.02 eV. Spin-polarized calculations were performed for all charge-states.

\subsection{\label{sec:defect_thermodynamics}Defect thermodynamics}
	
	The formation energy of a defect in charge-state $q$ is given by \cite{Zhang1991,Freysoldt2014}
	\begin{eqnarray}
		E^{\text{f}}(q) & = & E^{\text{tot}}_{\text{defect}}(q)-E^{\text{tot}}_{\text{bulk}}-\sum_{i}\Delta
		n_{i}\mu_{i} \nonumber \\
		& & + \ q(\varepsilon_{\text{VBM}}+\varepsilon_{\text{F}})+E^{\text{FNV}},
	\end{eqnarray}	
	where $E^{\text{tot}}$ are the electronic total energies, $\Delta n_{i}$ is the change in the number of atoms $i$ (Zn,O) with chemical potential $\mu_{i}$, $\varepsilon_{\text{F}}$ is the Fermi level relative to the bulk valence band (VB) maximum $\varepsilon_{\text{VBM}}$, and $E^{\text{FNV}}$ is an electrostatics-based finite-size correction term used to obtain the formation energy of the isolated defect from the finite-sized supercell calculation. We have employed the extended FNV correction scheme \cite{Freysoldt2009,Kumagai2014,Komsa2012}	
	\begin{equation}
		E^{\text{FNV}}=E^{\text{PC}}+q\Delta V^{\text{PC}}_{q/\text{bulk}|\text{far}},
	\end{equation}
	where $E^{\text{PC}}$ is the anisotropic Madelung energy for a periodic array of point charges immersed in a neutralizing background charge. The potential alignment term $\Delta V^{\text{PC}}_{q/\text{bulk}|\text{far}}$ is the difference between the defect-induced potential and point charge potential in a region far away from the defect	
	\begin{equation}
		\Delta V^{\text{PC}}_{q/\text{bulk}|\text{far}}=(V_{\text{defect},q}-V_{\text{bulk}})-V^{\text{PC}}_{q}. 
	\end{equation}	
	Since the atomic structure is allowed to relax when defects are introduced, the atomic site electrostatic potential is used as a potential marker, as discussed in detail by Kumagai \textit{et al.} \cite{Kumagai2014}. A small uncertainty in the alignment-like term arises due to the limited supercell size (less than 0.1 eV). The Madelung energy is estimated by an Ewald summation, and the macroscopic dielectric constant, valid for cubic systems, is replaced by a dielectric tensor. Both ion-clamped and ionic contributions to the dielectric tensor of the bulk system were calculated from self-consistent response of the system to a finite electric field \cite{Souza2002}, resulting in $\epsilon_{\perp}=7.19$ and $\epsilon_{\parallel}=8.23$. These values are lower than the experimental ones \cite{Ashkenov2003}, but closer than those obtained from density functional perturbation theory with the GGA-PBE functional \cite{Kumagai2014}.
	
	By varying the chemical potential, different experimental conditions can be explored. Upper and lower bounds are given by the stability of the phases that constitute the reservoir, which is expressed by the thermodynamic stability condition $\Delta H^{\text{f}}(\text{ZnO})=\mu_{\text{Zn}}+\mu_{\text{O}}$. The upper bound of $\mu_{\text{O}}$ (and thus the lower bound of $\mu_{\text{Zn}}$) is given by half the total energy of an O$_{2}$ molecule, and corresponds to O-rich conditions. Likewise, the lower limit of $\mu_{\text{O}}$ (and the upper limit of $\mu_{\text{Zn}}$) is given by the reduction of ZnO to metallic Zn, corresponding to Zn-rich conditions \cite{Janotti2007}. The hybrid functional yields a ZnO heat of formation of $\Delta H^{\text{f}}(\text{ZnO}) = -3.49$ eV, which is close to the experimental value of $-3.61$ eV \cite{Parks1927}.
	
	From the calculated defect formation energies, thermodynamic charge-state transition levels are given by the Fermi level position for which the formation energy of the defect in two charge-states $q_{1}$ and $q_{2}$ is equal, i.e., \cite{Freysoldt2014}
	\begin{equation}\label{eq:transition-level}
		\varepsilon(q_{1}/q_{2})=\frac{E^{\text{f}}(q_{1};\varepsilon_{\text{F}}=0)-E^{\text{f}}(q_{2};\varepsilon_{\text{F}}=0)}{q_{2}-q_{1}}.
	\end{equation}

\subsection{\label{sec:defect_luminescence}Defect luminescence}
	
\begin{figure}[!htb]
		\includegraphics[width=0.9\columnwidth]{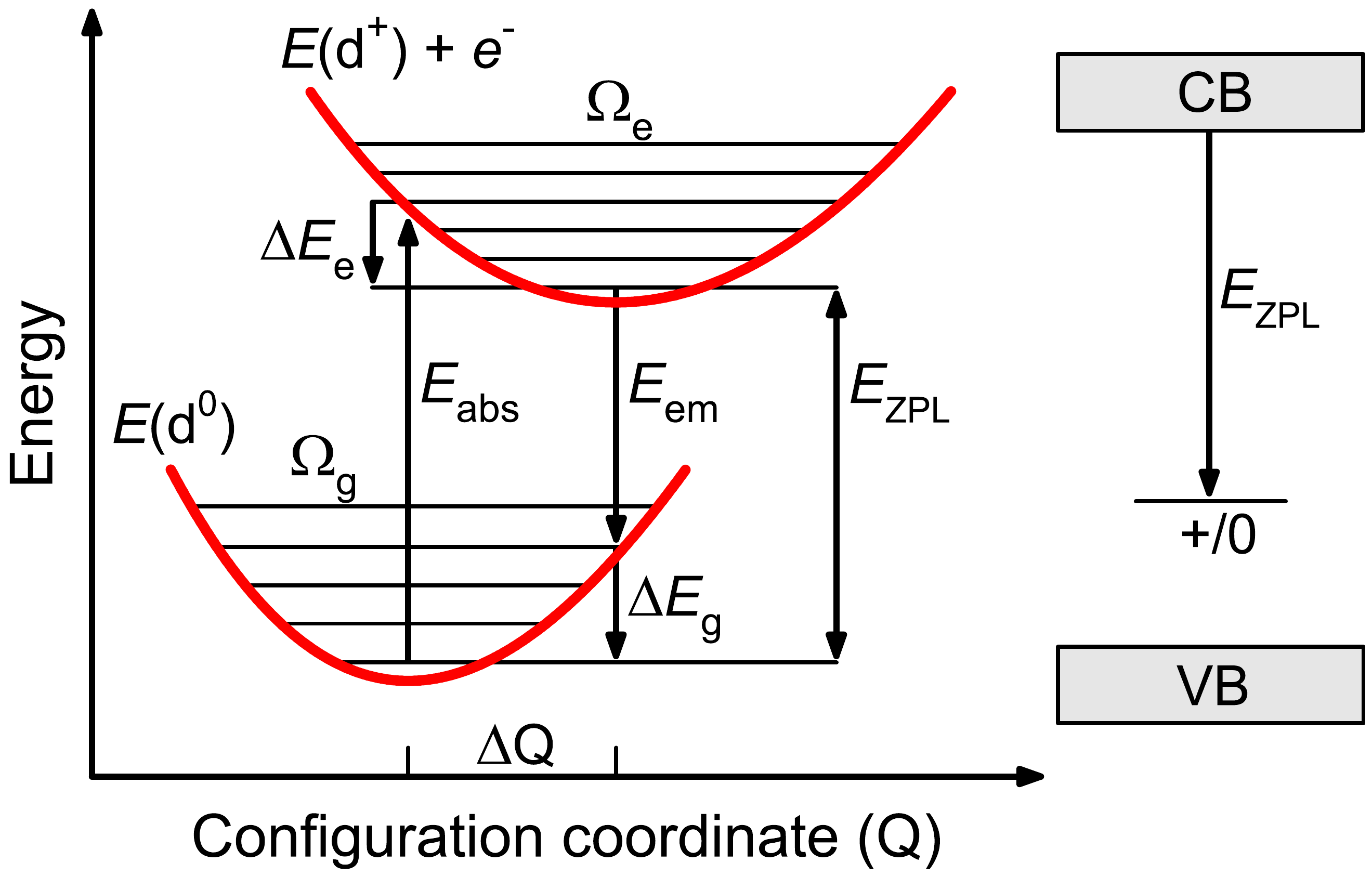}
		\caption{Configuration coordinate diagram illustrating vibronic transitions between the neutral and singly positive charge-state of a defect. $E_{\text{ZPL}}$ gives the thermodynamic charge-state transition level relative to the conduction band (CB) minimum (shown in the band diagram on the right). The horizontal lines in each normal mode are vibrational energy levels, and the probability of any given vibronic transition is proportional to the vibrational wave function overlap between the initial level and final level.}
		\label{fig:CC_diagram}
	\end{figure}

	Defect luminescence lineshapes were calculated by using the methodology described in Ref. \onlinecite{Alkauskas2012}, wherein the multidimensional vibrational problem is mapped onto an effective one-dimensional configuration coordinate diagram \cite{Alkauskas2012,Stoneham} (Fig. \ref{fig:CC_diagram}).
	The parameters that enter the model are the effective phonon frequencies $\Omega_{\text{g/e}}$, the zero phonon line $E_{\text{ZPL}}$ and the configuration coordinate $\Delta\text{Q}$ which is defined as
	\begin{equation}\label{Q}
		(\Delta\text{Q})^{2}=M\Delta R^{2}=\sum_{i\alpha}m_{\alpha}(R_{\text{e},i\alpha}-R_{\text{g},i\alpha})^{2}.
	\end{equation}	
	Here, $M$ is the effective modal mass in atomic units and $\Delta R$ is the magnitude of the displacement in \AA \ for all atoms $\alpha$ in the supercell ($i=\{x,y,z\}$). Thus, the configuration coordinate represents the collective motion of all atoms in the supercell between the different charge-states, meaning that the various individual vibrational modes are replaced by a single effective mode. All parameters are obtained from the hybrid DFT calculations by using finite differences.
	
	Huang-Rhys (HR) factors \cite{Huang1950,Stoneham} describe the average number of phonons that are involved in a transition, and can be expressed as $S_{\text{g}} = \Delta E_{\text{g}}/\hbar\Omega_{\text{g}}$ for emission. $\Delta E_{\text{g}}$ is the relaxation energy, often referred to as the Franck-Condon shift. The effective one-dimensional configuration coordinate model is a good approximation for broad luminescence bands with strong electron-phonon coupling ($S\gg 1$), as demonstrated in Refs. \onlinecite{Alkauskas2012,Alkauskas2016}.

\subsection{\label{sec:si_error}The self-interaction error}
	
	While the value of the fraction of HF exchange used (37.5\%) reproduces the lattice parameters and bulk band gap of ZnO, this does not necessarily mean that the defect states of $V_{\text{Zn}}$ are described correctly \cite{Ivady2013}. In order to elaborate on possible over-localization, all charge-states were also calculated using the original HSE06 functional (25\% HF exchange). However, the results remained qualitatively unaffected regarding the localization of holes. The energy positions of the polaronic Kohn-Sham (KS) states, relative to the average electrostatic potential, were almost unchanged. Moreover, we found the so-called non-Koopmans' energy for $V_{\text{Zn}}^{0}$, defined in Refs. \onlinecite{Ivady2013,Lany2009} as	
	\begin{equation}\label{eq:NK}
		E_{\text{NK}}=\varepsilon(N)-E_{\text{A}}=\varepsilon(N)-(E(N+1)-E(N)),
	\end{equation}	
	to be small (0.12  eV). Here, $\varepsilon(N)$ is the KS quasiparticle energy of the polaronic state and $E_{\text{A}}$ is the electron addition energy of the system, i.e., the difference in total energy between the $(N+1)$- and $N$-electron system, keeping the ions fixed to their $N$-electron ground-state positions. The electrostatic finite-size correction was applied only to $E(N+1)$, using the ion-clamped dielectric tensor, since the two remaining terms are for the neutral defect. $E_{\text{NK}}$ may still contain a small finite-size error. We conclude, however, that the self-interaction error is small in the calculations, and, importantly, the qualitative results do not hinge on the specific value used for the exchange parameter.

\section{\label{sec:results}Results and Discussion}
	
	In wurtzite ZnO, four O ions form a tetrahedron around every Zn ion and vice versa. In these tetrahedra, we shall refer to the ions in the three corners of the basal plane as azimuthal ions, while the ion in the fourth corner will be referred to as the axial ion. When a Zn vacancy is formed, four Zn--O bonds are broken. The dangling O 2\textit{p} bonds that remain are partially filled by six electrons, and can accommodate two more. This simple chemical picture dictates that $V_{\text{Zn}}$ acts as a double acceptor. However, it can also trap holes in polaronic states, as will be shown in Sections \ref{sec:thermodynamics} and \ref{sec:polaron}.

\subsection{\label{sec:thermodynamics}Thermodynamics of the Zn vacancy}
	
\begin{figure}[!htb]
		\includegraphics[width=1.0\columnwidth]{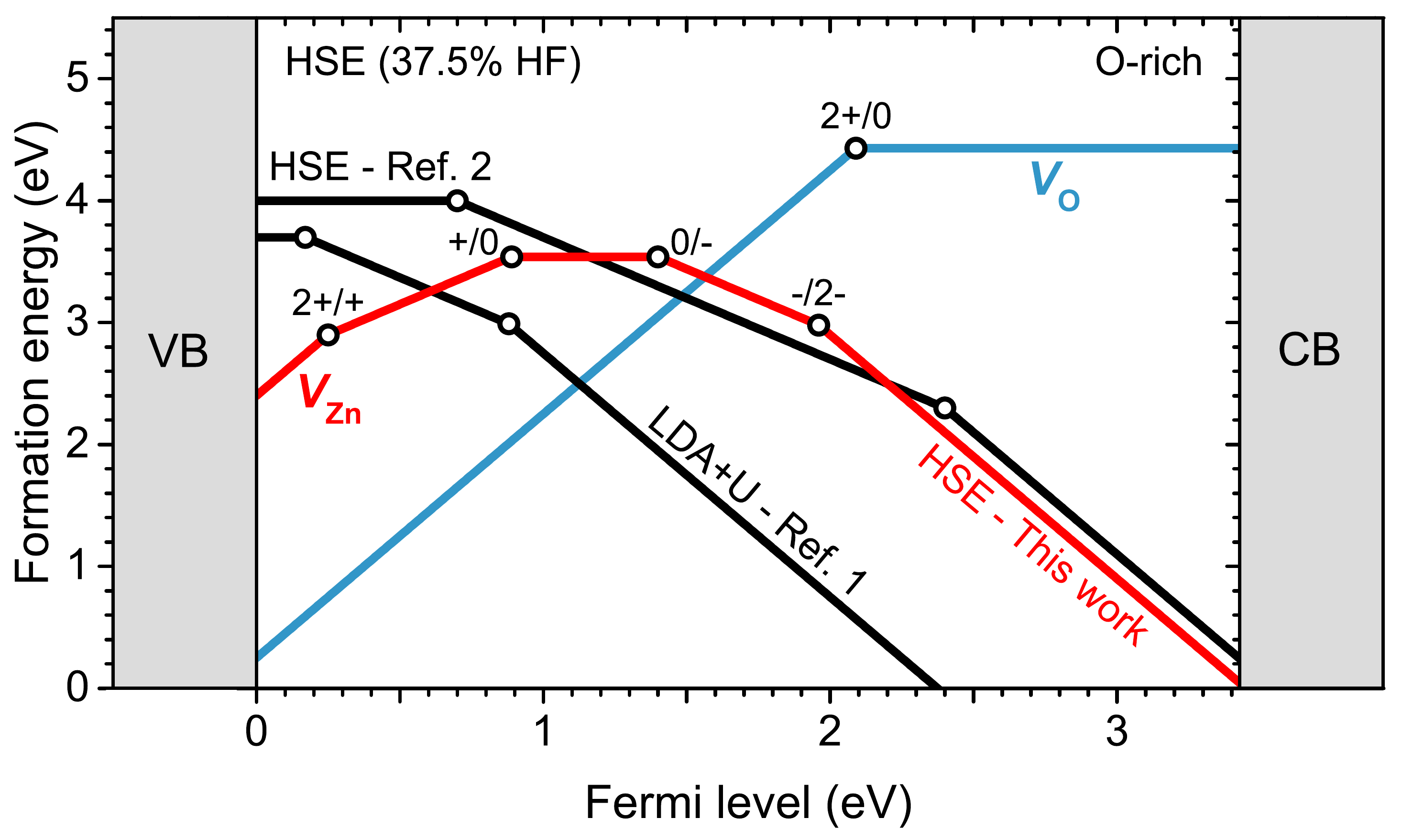}
		\caption{The formation energy of $V_{\text{Zn}}$ (red line) as a function of the Fermi level position (from the VB maximum to the CB minimum), under O-rich conditions. Thermodynamic charge-state transition levels are labeled. Results from two previous DFT studies are included for comparison (black lines), and the formation energy of $V_{\text{O}}$ (blue line) is included as a benchmark (see text).}
		\label{fig:FE-V_Zn}
	\end{figure}

	Fig. \ref{fig:FE-V_Zn} shows the formation energy of $V_{\text{Zn}}$ as a function of the Fermi level position under O-rich conditions. The formation energy approaches $\sim$0.2 eV near the CB minimum, which indicates that $V_{\text{Zn}}$ should be the dominating intrinsic acceptor in \textit{n}-type ZnO. This is in agreement with positron annihilation spectroscopy (PAS) measurements \cite{Tuomisto2003} and previous DFT studies \cite{Janotti2007,Oba2008}. However, the majority of previous DFT studies have only included acceptor charge-states of $V_{\text{Zn}}$ (two examples are shown in Fig. \ref{fig:FE-V_Zn}). As shown previously \cite{Bjorheim2012,Lany2009}, $V_{\text{Zn}}$ can display positive charge-states as well. Indeed, we find both the + and 2+ state, with thermodynamic transitions located at 0.25 (2+/+), 0.89 (+/0), 1.40 (0/-) and 1.96 eV (-/2-) above the VB maximum (note that the transitions are approximately equidistant). The emergence of both positive and negative charge-states means that $V_{\text{Zn}}$ is an amphoteric defect. Although the formation energy is rather high in \textit{p}-type material, $V_{\text{Zn}}$ is predicted to act as a compensating donor in a frozen-in, out-of-equilibrium scenario.
	
	One might question why the previous hybrid calculations by Oba \textit{et al}. \cite{Oba2008} deviate so much from our results. The main reason for this is the fact that spin-polarization was taken into account for $V_{\text{Zn}}^{-}$ only in Ref. \onlinecite{Oba2008} (due to computational constraints), but also because a plane-wave energy cutoff of 300 eV and a $\Gamma$-only \textit{k}-point sampling were used. We elaborate on this in Section \ref{sec:polaron}.
	
	The calculated position of the (0/-) transition agrees well with photo-EPR data; Evans \textit{et al.}\cite{Evans2008} inferred that the threshold energy to excite an electron from $V_{\text{Zn}}^{-}$ to the CB (to observe the EPR signal of $V_{\text{Zn}}^{0}$) is $\sim$2.5 eV. We obtain an absorption energy of 2.64 eV for $V_{\text{Zn}}^{-}$, as shown in Fig. \ref{fig:CC-V_Zn} (b). This is somewhat higher than the photo-EPR data, but the onset shifts down due to vibrational broadening. This can be shown by simulating the absorption profile, as demonstrated in Ref. \onlinecite{Alkauskas2016}.
	
	The O vacancy ($V_{\text{O}}$) is also included in Fig. \ref{fig:FE-V_Zn} for comparison, since the (2+/0) transition level of this defect has become a benchmark case for defects in ZnO \cite{Alkauskas2011}. In fact, the defect state of $V_{\text{O}}$ is fairly well described by a wide range of different functionals. While there is a large spread in the reported thermodynamic charge-state transition levels relative to the VB maximum, the agreement becomes decent when they are aligned to a common reference level, such as the average electrostatic potential \cite{Alkauskas2011}. We obtain 2.1 eV above the VB maximum for the (2+/0) transition, which is in good agreement with previous calculations based on hybrid functionals \cite{Oba2008,Alkauskas2011,Clark2010}. The defect wave function of $V_{\text{Zn}}$, on the other hand, is poorly described at the (semi)local level, implying that the energy positions of the  thermodynamic charge-state transition levels depend strongly on the choice of functional.

\subsection{\label{sec:polaron}Polaronic hole localization}
	
\begin{figure}[!htb]
		\includegraphics[width=1.0\columnwidth]{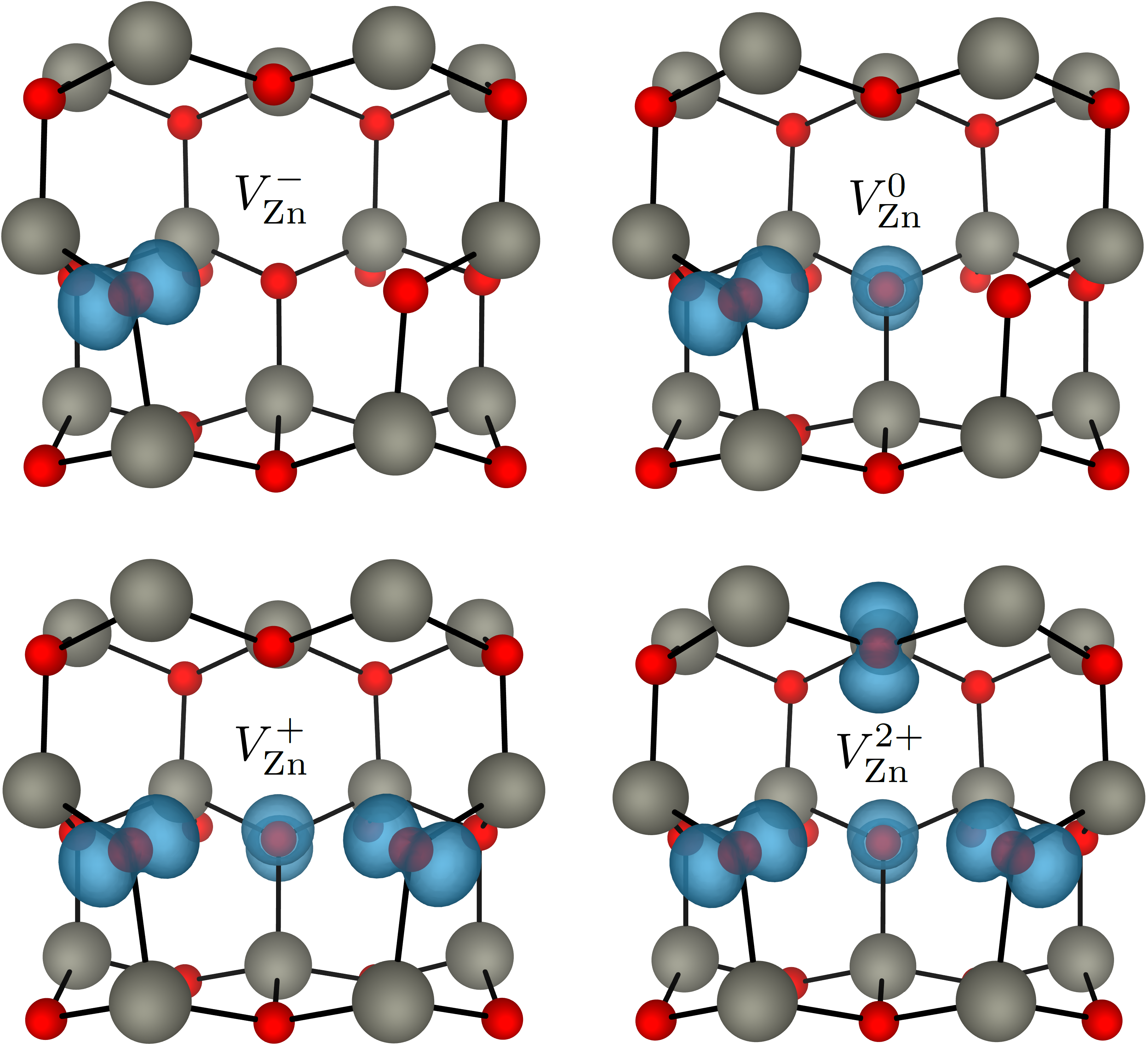}
		\caption{Probability density for holes occupying polaronic KS eigenstates in $V_{\text{Zn}}$ (blue). When the HSE functional is used, holes lock onto separate O ions (red). When a semilocal functional is used, the holes delocalize over several O ions due to the self-interaction error. The wavefunctions have a distinct O 2\textit{p} orbital character. The probability density isosurface was set to 0.02 $r_{\text{Bohr}}^{-3}$.}
		\label{fig:polaron-V_Zn}
	\end{figure}

	The emergence of the positive charge-states of $V_{\text{Zn}}$ can be understood by taking a closer look at its electronic and atomic structure. First $V_{\text{Zn}}^{2-}$ is considered, where the O 2\textit{p} dangling bond states are completely filled with electrons. By examining the \textit{spd}- and site-projected wave function character of each KS state, one can deduce that the dangling bonds introduce three states in the band gap close to the VB maximum, and one resonant with the VB. When one electron is removed, the spin-degeneracy of the resulting half-filled defect state is broken, and the empty state moves deep into the band gap. As more electrons are removed, additional empty states exhibiting polaronic nature appear deep in the band gap, until all four dangling bond states are half-filled ($V_{\text{Zn}}^{2+}$). The probability density of the empty defect states, shown in Fig. \ref{fig:polaron-V_Zn}, illustrates that each hole localizes onto a single nearest-neighbor O$^{2-}$ ion in the form of a small hole polaron. This spontaneous localization of holes is accompanied by a distinct outward relaxation of the O$^{-}$ ion, which further moves the polaronic state into the band gap, lowering the total defect energy. The azimuthal O$^{-}$ ions with trapped holes move away from the vacancy by approximately 14\% of the bulk Zn--O bond length, which is about twice as far as the azimuthal O$^{2-}$ ions without trapped holes. This behavior (hole localization with local lattice distortion) is common for many oxide semiconductors \cite{Varley2012,Lyons2014}.

	As pointed out by Janotti \textit{et al.} \cite{Janotti2007}, the (semi)local functionals are unable to describe the Zn vacancy (and thus fail to stabilize $V_{\text{Zn}}^{+}$ and $V_{\text{Zn}}^{2+}$). This is because of the self-interaction error; a lower total energy occurs by dividing the hole between multiple O ions. Lany and Zunger \cite{Lany2009} removed this delocalization bias by using a hole-state potential to enforce fulfillment of the generalized Koopmans' condition, ensuring a linear behavior of the total energy and a constant behavior of the highest-occupied single-particle level with respect to fractional occupation \cite{Lany2009,Ivady2013}. This correction stabilized $V_{\text{Zn}}^{+}$ and $V_{\text{Zn}}^{2+}$, but it does not remedy the severe band gap underestimation of GGA ($E_{\text{g}}=0.73$ eV), leading to an ambiguity in the energy position of the thermodynamic charge-state transition levels with respect to the band edges. It must be noted, however, that the overall result of Lany and Zunger is in good agreement with our result.
	
	Incorporating a fraction of exact exchange, hybrid functionals cancel (at least in part) the self-interaction error, and provide accurate band gaps. However, previous hybrid DFT studies employing the HSE, PBE0 and sX functionals did still not reveal the positive charge-states of $V_{\text{Zn}}$ \cite{Oba2008,Clark2010}. Here, we demonstrate that initial symmetry-breaking operations, like moving the O ions slightly or specifying their initial magnetic moment, are a prerequisite to obtain localization of the holes onto single O ions. It is also crucial that the ions are relaxed with the hybrid functional, and that spin-polarized calculations are performed for all charge-states. Otherwise, the ground-state will not be obtained; the holes may instead delocalize over more than one O ion, with no polaronic effects. By breaking the symmetry, all metastable localized hole configurations were investigated. The azimuthal configuration of holes, shown in Fig. \ref{fig:polaron-V_Zn}, was found to be the most stable one---in agreement with EPR data \cite{Galland1970,Galland1974,Kappers2008,Evans2008}. In addition, a seperation of 3.69 \AA \ between the two O$^{-}$ ions with trapped holes in $V_{\text{Zn}}^{0}$ was obtained, which is close to the 3.75 \AA \ inferred from EPR measurements \cite{Galland1974,Evans2008}. Finally, we find that the high-spin configuration of $V_{\text{Zn}}$ is energetically preferred, which means that \textit{S}=1/2 for $V_{\text{Zn}}^{-}$, \textit{S}=1 for $V_{\text{Zn}}^{0}$, \textit{S}=3/2 for $V_{\text{Zn}}^{+}$ and \textit{S}=2 for $V_{\text{Zn}}^{2+}$.
	
	Polaronic hole localization is not unique for $V_{\text{Zn}}$ in ZnO. While lattice deformation alone is not sufficient to induce hole localization \cite{Varley2012}, polarons can form when an acceptor-like defect exists at a neighboring Zn site \cite{Lyons2014}. Indeed, substitutional Group-I impurities (Li$_{\text{Zn}}$, Na$_{\text{Zn}}$) exhibit the same tendency to stabilize anion trapped hole polarons \cite{Du2009,Carvalho2009,Bjorheim2012}. Moreover, anion site substitutional impurities can lead to deep atomic-like localized states \cite{Lyons2014}. These effects, in combination with the very low position of the ZnO VB on an absolute energy scale and the heavy hole effective masses, render \textit{p}-type doping of ZnO challenging at equilibrium conditions.

\subsection{\label{sec:energetics_model}A simple model for the Zn vacancy energetics}
	
	Here, a simple model to explain the energetics of $V_{\text{Zn}}$, with regard to the approximately equidistant charge-state transition levels, is presented. First, assume the Fermi level position at the VB maximum ($\varepsilon_{\text{F}}=0$ eV in Fig. \ref{fig:FE-V_Zn}), and as a starting point consider $V_{\text{Zn}}^{2-}$. Adding a hole to the defect lowers the formation energy by $\varepsilon_{0}$, which includes the polaron formation energy. Subsequently	
	\begin{equation}\label{key}
		E^{\text{f}}(V_{\text{Zn}}^{-})=E^{\text{f}}(V_{\text{Zn}}^{2-}) - \varepsilon_{0}.
	\end{equation}
	According to Eq. (\ref{eq:transition-level}), this translates into the $\varepsilon$(-/2-) charge-state transition level being located at
	\begin{equation}\label{key}
		\varepsilon\text{(-/2-)}=\varepsilon_{0}
	\end{equation}	
	above the VB maximum. Adding another hole lowers the formation energy of $V_{\text{Zn}}^{-}$ by $\varepsilon_{0}-U$, where $U$ is the hole-hole repulsion energy, that is
	\begin{equation}\label{key}
		\varepsilon\text{(0/-)}=\varepsilon_{0}-U.
	\end{equation}
	Adding a third hole lowers the energy even less, as now the hole is repelled by two other holes:
	\begin{equation}\label{key}
		\varepsilon\text{(+/0)}=\varepsilon_{0}-2U.
	\end{equation}
	Similarly for the fourth hole
	\begin{equation}\label{key}
		\varepsilon\text{(2+/+)}=\varepsilon_{0}-3U.
	\end{equation}
	This simple model explains why the $V_{\text{Zn}}$ charge-state transition levels are approximately equidistant, since
	\begin{eqnarray}\label{key}
		\varepsilon\text{(-/2-)}-\varepsilon\text{(0/-)} & \cong & U \\
		\varepsilon\text{(0/-)}-\varepsilon\text{(+/0)} & \cong & U \\
		\varepsilon\text{(+/0)}-\varepsilon\text{(2+/+)} & \cong & U.
	\end{eqnarray}
	Of course, this is only approximately true. The hole-hole repulsion $U$ corresponds to the so-called Hubbard correlation energy \cite{Hubbard1963}. By taking the average of Eq. (12--14), the value can be derived as $U\simeq0.57$ eV. Additionally, the fact that the in-plane configuration of holes is energetically preferred explains why the separation between the (2+/+) and (+/0) level is somewhat larger than the other two; the fourth hole must localize on the remaining axial O ion (lower hole addition energy). Taking this into account, i.e., considering only Eq. (12) and (13), the hole-hole repulsion energy becomes $U\simeq0.53$ eV. 
	
	These considerations help rationalize our findings with reference to the energetics of $V_{\text{Zn}}$. In fact, by inspecting the results of Ref. \onlinecite{Lyons2015}, we suggest that a similar model applies to the Ga vacancy ($V_{\text{Ga}}$) in GaN, which can also trap up to four holes at the nearest neighbor N ions.

\subsection{\label{sec:luminescence}Optical properties of the Zn vacancy}

	\begin{figure*}[htb!]
		\includegraphics[width=\textwidth]{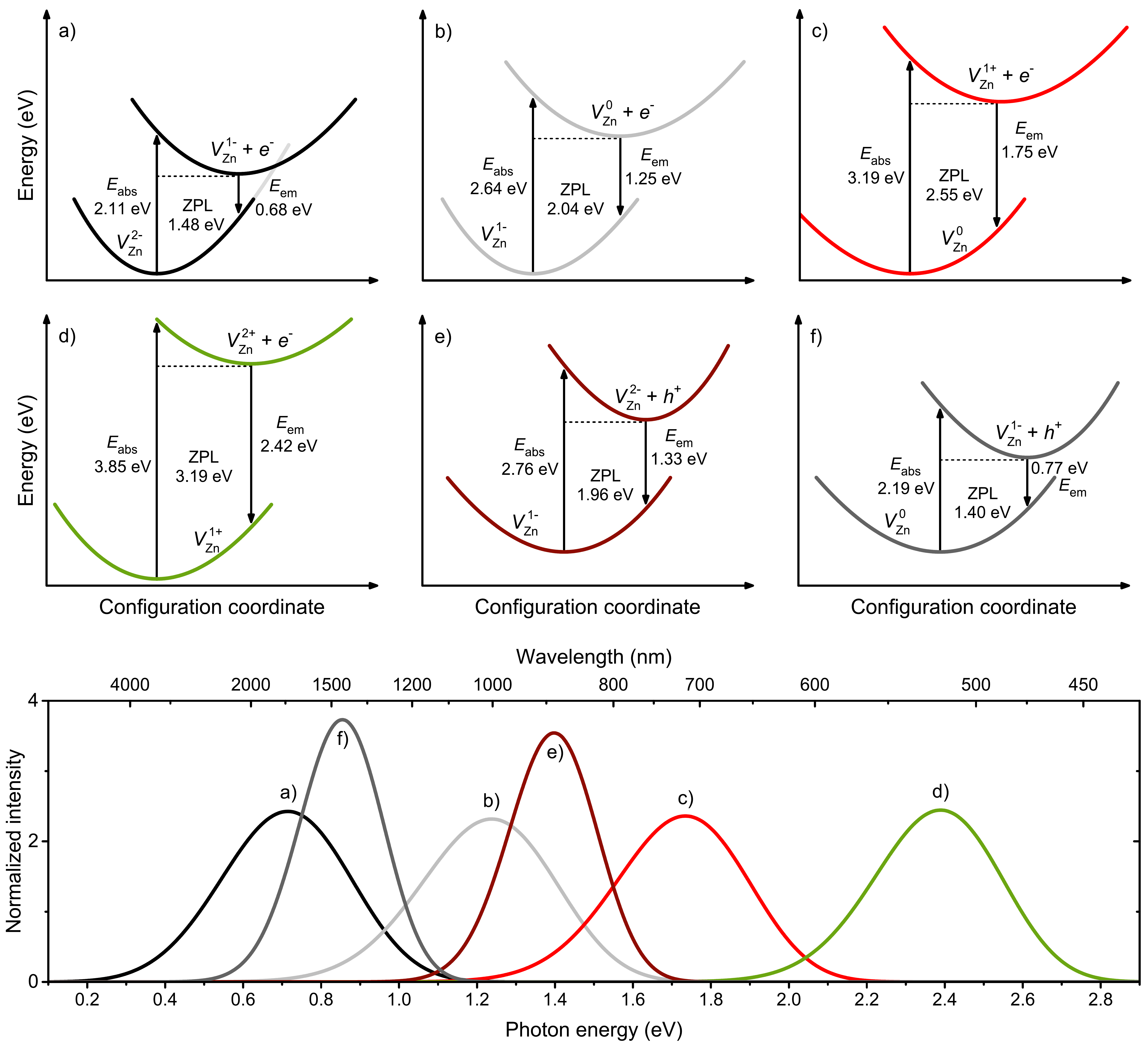}
		\caption{Configuration coordinate diagrams and luminescence lines for optical transitions involving the Zn vacancy. The respective transitions are color-matched and labeled from a) to f) in the configuration coordinate diagrams and luminescence spectrum. From left to right, the peak position (PP) of the luminescence bands are at 0.71, 0.85, 1.24, 1.40, 1.74 and 2.39 eV.}
		\label{fig:CC-V_Zn}
	\end{figure*}
	
	\begin{table*}[htb!]
		\caption{Effective parameters for the calculated Zn vacancy luminescence transitions in ZnO; total mass-weighted distortion ($\Delta\text{Q}$), effective ground- and excited-state normal mode frequencies ($\hbar\Omega_{\text{g/e}}$), zero phonon line energy ($E_{\text{ZPL}}$), peak position (PP), full width at half maximum of the luminescence band (FWHM), ground- and excited-state Huang-Rhys factors ($S_{\text{g/e}}$). \label{table:effective-parameters}}
		\begin{ruledtabular}
			\begin{tabular}{cccccccccc}
				Transition & $\Delta\text{Q}$ (amu$^{1/2}$\AA) & $\hbar\Omega_{\text{g}}$ (meV) & $\hbar\Omega_{\text{e}}$ (meV) & $E_{\text{ZPL}}$ (eV) & PP (eV) & FWHM (eV) & $S_{\text{g}}$ & $S_{\text{e}}$ \\
				\hline
				a) $\mathrm{\mathit{V}_{Zn}^{-} + e^{-} = \mathit{V}_{Zn}^{2-}}$ & 2.60 & 34 & 27 & 1.48 & 0.71 & 0.39 & 23.4 & 23.1 \\
				b) $\mathrm{\mathit{V}_{Zn}^{0} + e^{-} = \mathit{V}_{Zn}^{-}}$ & 2.77 & 31 & 25 & 2.04 & 1.24 & 0.41 & 25.2 & 24.7 \\
				c) $\mathrm{\mathit{V}_{Zn}^{+} + e^{-} = \mathit{V}_{Zn}^{0}}$ & 2.90 & 30 & 25 & 2.55 & 1.74 & 0.40 & 26.8 & 26.4 \\
				d) $\mathrm{\mathit{V}_{Zn}^{2+} + e^{-} = \mathit{V}_{Zn}^{+}}$ & 3.00 & 28 & 24 & 3.19 & 2.39 & 0.38 & 27.2 & 27.5 \\
				e) $\mathrm{\mathit{V}_{Zn}^{2-} + h^{+} = \mathit{V}_{Zn}^{-}}$ & 2.60 & 27 & 34 & 1.96 & 1.40 & 0.27 & 23.2 & 23.5 \\
				f) $\mathrm{\mathit{V}_{Zn}^{-} + h^{+} = \mathit{V}_{Zn}^{0}}$ & 2.77 & 25 & 31 & 1.40 & 0.85 & 0.25 & 24.7 & 25.2 \\
			\end{tabular}
		\end{ruledtabular}
	\end{table*}
	
	Optical transitions involving $V_{\text{Zn}}$ and the band edges have been investigated. The configuration coordinate diagrams and calculated lineshapes and positions are shown in Fig. \ref{fig:CC-V_Zn}, and all the effective parameters for the various transitions are given in Table \ref{table:effective-parameters}. We find effective normal mode frequencies $\hbar\Omega_{\text{g/e}}$ between 24--34 meV, total mass-weighted distortions between 2.6--3.0 amu$^{1/2}$\AA \ and large HR factors between 23--28, resulting in broad luminescence lines. This is expected for optical transitions involving polaronic acceptors like $V_{\text{Zn}}$, because of the sizeable changes in the atomic geometry between different charge-states \cite{Lyons2014,Alkauskas2012}. Indeed, the main contribution to $\Delta\text{Q}$ comes from the four nearest Zn ions to the O$^{-}$ ions with trapped holes.
	
	Considering ZnO as primarily an \textit{n}-type material, optical transitions involving $V_{\text{Zn}}^{+}$ and $V_{\text{Zn}}^{2+}$ require $V_{\text{Zn}}^{2-}$ to rapidly trap three and four holes, respectively. This is perhaps an unlikely scenario (unless the concentration of photogenerated holes is extremely high). Accordingly, we restrict primarily our following discussion to transitions involving $V_{\text{Zn}}^{2-}$, $V_{\text{Zn}}^{-}$ and $V_{\text{Zn}}^{0}$.
	
	Capture of an electron located at the CB minimum by $V_{\text{Zn}}^{-}$ results in a broad luminescence lineshape peaking at an energy of 0.71 eV. Note, however, that the excited- and ground-state normal modes overlap close to the minimum of the excited state (illustrated in Fig. \ref{fig:CC-V_Zn} (a)). The energy barrier from the minimum of the excited state up to the point of intersection is only 90 meV, implying that the transition is likely to be nonradiative. Hole capture by $V_{\text{Zn}}^{0}$ and V$_{\text{Zn}}^{+}$ is expected to be nonradiative for the same reason, and have been omitted from Fig. \ref{fig:CC-V_Zn}. In fact, in the latter case, the effective normal modes intersect at $\text{Q}<\Delta\text{Q}$, i.e., before the minimum of the excited state is reached. In contrast, electron capture by $V_{\text{Zn}}^{0}$ will have both a radiative and nonradiative component, since the energy barrier is 0.48 eV. The resulting luminescence lineshape peaks at 1.24 eV.

	Capture of a hole located at the VB maximum by $V_{\text{Zn}}^{2-}$ results in a somewhat narrower luminescence band peaking at 1.40 eV (Fig. \ref{fig:CC-V_Zn} (e)). Strictly speaking, since the VB in ZnO and the defect states of $V_{\text{Zn}}$ both have O 2\textit{p} character, this transition should be forbidden. However, just like for hole capture by $V_{\text{Ga}}^{3-}$ in GaN \cite{Lyons2015}, the transition may be allowed because of the strong polaronic relaxation. Nevertheless, this transition has to compete with shallower negatively charged acceptors like Li$_{\text{Zn}}^{-}$, which captures holes nonradiatively in an efficient manner \cite{Alkauskas2014a}. Following this logic, the luminescence might be weak in reality, depending on the purity of the sample and the concentration of $V_{\text{Zn}}$.

	Finally, it should be pointed out that the simulation results in Fig. \ref{fig:CC-V_Zn}, revealing prevalent $V_{\text{Zn}}$-related luminescence at low energies close to the infrared region in \textit{n}-type material, are consistent with recent experimental data by Dong \textit{et al}. \cite{Dong2010} and Knutsen \textit{et al}. \cite{Knutsen2012}. Through a combination of cathodoluminescence and PAS measurements, it was found in Ref. \onlinecite{Dong2010} that the emission from small $V_\text{Zn}$ clusters peaks at $\sim$1.9 eV and shifts to lower energies with decreasing cluster size. Similarly, using samples irradiated with electrons having energies below and above the threshold for displacement of Zn atoms, as well as samples annealed in Zn-rich and O-rich ambients, Knutsen \textit{et al}. \cite{Knutsen2012} demonstrated that the luminescence in the near infrared region arises from $V_{\text{Zn}}$, or defects containing $V_{\text{Zn}}$. Especially in the case where $V_{\text{Zn}}$ forms a complex with a donor, e.g., $V_{\text{O}}$ \cite{Johansen2016}, the most negative thermodynamic charge-state transition levels would be passivated by the donor electrons. Hence, one could speculate that a transition similar to c) in Fig. \ref{fig:CC-V_Zn}, peaking in the 1.6--1.9 eV range, would prevail for such a complex, consistent with the experimental data \cite{Dong2010,Knutsen2012}.

\section{\label{sec:conclusion}Conclusion}
	
	Based on the present calculations, we conclude that $V_\text{Zn}$ in ZnO is a deep polaronic acceptor that can bind a localized hole on each of its four nearest-neighbor O ions. The distinct outward relaxation of these O ions is a key feature of the polaronic nature of $V_\text{Zn}$---in agreement with experimental EPR data \cite{Galland1970,Galland1974,Kappers2008,Evans2008}.
	
	By employing a one-dimensional configuration coordinate model \cite{Alkauskas2012}, luminescence positions and lineshapes from $V_\text{Zn}$ were simulated. In contrast to what has been previously suggested \cite{Wang2015,Zhao2006,Wang2011,Wang2009,Janotti2007}, the present results show that the isolated $V_\text{Zn}$ is unlikely to be a major source of luminescence in the visible range for \textit{n}-type material. All transitions involving $V_{\text{Zn}}^{2-}$, $V_{\text{Zn}}^{-}$ and $V_{\text{Zn}}^{0}$ are nonradiative and/or lead to luminescence lineshapes that are very low in energy (near-infrared region). Electron capture by $V_{\text{Zn}}^{+}$ and $V_{\text{Zn}}^{2+}$ leads to red and green luminescence, respectively, but these transitions are unlikely to occur in \textit{n}-type material unless the concentration of photogenerated holes is extremely high. These results are consistent with recent experimental data \cite{Dong2010,Knutsen2012}.
	
	The luminescence lines arising from $V_\text{Zn}$ are broad. This is because of the large relaxation associated with hole capture by one of the nearest-neighbor O ions, i.e., the strong electron-phonon coupling. This highlights the important role of hybrid functionals, which, unlike (semi)local functionals, are able to predict charge localization associated with local lattice distortions around defects \cite{Freysoldt2014}.

\begin{acknowledgments}
	
	We wish to thank Prof. F. Oba for helpful discussions. Financial support was kindly provided by the Research Council of Norway and University of Oslo through the frontier research project FUNDAMeNT (no. 251131, 	FriPro ToppForsk-program). K. M. Johansen would like to thank the Research Council of Norway for support to the DYNAZOx project (no. 221992). A.A. was supported by Marie Sk\l{}odowska-Curie Action of the European Union (project \textsc{Nitride}-SRH, Grant No. 657054). UNINETT Sigma2 and the Department for Research Computing at the University of Oslo are acknowledged for providing computational resources and support under Projects No. NN4604K, NN9180K and NN9136K.
	
\end{acknowledgments}
	
\bibliography{Main-text-with-figures}	

\end{document}